\documentclass[preprint]{JHEP3}

\catcode`@=11 \@addtoreset{equation}{section} \catcode`@=12

\title{Exact Effective Superpotential for $SO(N_c)$ Gauge Theory
with $N_f$ Flavors }

\author{Pravina Borhade and P. Ramadevi\\Department of Physics, Indian Institute of Technology Bombay, Mumbai 400 076, India\\ E-mail: \email{pravina@phy.iitb.ac.in}, \email{ramadevi@phy.iitb.ac.in}
}

\preprint{\hepth{0605100}}

\abstract{Motivated by the duality conjecture of
Dijkgraaf and Vafa between supersymmetric gauge
theories and matrix models, we derive the effective
superpotential of ${\cal N}=1$ supersymmetric gauge theory 
with gauge group $SO(N_c)$ and arbitrary tree level polynomial superpotential 
of one chiral superfield in the adjoint representation and $N_f$ fundamental
matter multiplets.} 


\begin{document}

\section{Introduction}
Large $N$ topological duality relating $U(N)$ Chern-Simons gauge theory 
on $S^3$ to $A$-model topological string \cite{gv1} and its embedding 
in the superstring context \cite{vafa1} has led to interesting 
interconnections between geometry of Calabi-Yau 
three-folds ($CY_3$) and ${\cal N}=1$ supersymmetric gauge theories.
Strong coupling dynamics of supersymmetric gauge theories
can be studied within the superstring duality \cite{vafa1} by
geometrically engineering $D$-branes. Using the
geometric considerations of dualities in IIB string theory,
Cachazo et al \cite{civ1} have obtained low energy effective 
superpotential for a class of $CY_3$ geometries whose
singular limit is given by
\begin{equation}
W'(x)^2+y^2+z^2+w^2=0~, \label {singu}
\end{equation}
where $W(x)$ is a polynomial of degree $n+1$. 
In fact, the low energy effective superpotential corresponds to a  
${\cal N}=1$ supersymmetric $U(N)$ Yang-Mills with adjoint scalar $\Phi$ 
and tree level superpotential $W_{tree}(\Phi)=\sum_{k=1}^{n+1} (g_k/k)
Tr \Phi^k$.

The mirror version of the large $N$ topological duality conjecture\cite{gv1}
was considered in ref. \cite{dv1} relating topological $B$ strings
on the $CY_3$ geometries \cite{civ1} to matrix models. The potential of the 
matrix model $W(\Phi)=(1 / g_s) W_{tree}(\Phi)$  where $\Phi$ 
denotes a hermitean matrix.  Further, Dijkgraaf-Vafa have 
conjectured that the low-energy effective superpotential can be 
obtained from the planar limit of these matrix models \cite{dv1,dv2,dv3}. 
The Dijkgraaf-Vafa conjecture was later on proved by various methods: 
(i) by factorization of Seiberg-Witten curves \cite {ferr1}, 
(ii) using perturbative field theory arguments 
\cite {dglvz1} and (iii) generalized Konishi anomaly approach \cite {cdsw1}.

The extension of topological string duality relating
Chern-Simons theory with $SO/Sp$ gauge groups to $A$-model
closed string on an orientifold of the resolved conifold
was studied by Sinha-Vafa \cite {sv1}.
Generalizing the geometric procedure considered for $U(N)$ \cite {civ1}, 
the effective superpotential for ${\cal N}=1$ supersymmetric 
theories with $SO/Sp$ gauge groups with 
$W_{tree}= \sum_k (g_{2k}/  2k) Tr \Phi^{2k}$ where $\Phi$ is
adjoint scalar superfield  were derived for the
orientifolds of the $CY_3$ geometries \cite {eot1}. 
These effective superpotentials have also been
computed within perturbative gauge theory \cite {iho1},
using  matrix model techniques in \cite {achkr1} and using
the factorization property of ${\cal N}=2$ Seiberg-Witten curves
\cite {jo1}. Related works involving second rank tensor matter fields
have been considered in refs.\cite{ikrsv1,argurio1,ac1}.

So far, the effective superpotential computation involved
${\cal N}=1$ supersymmetric gauge theories with either adjoint matter
or second rank tensor matter. 
The inclusion of matter transforming in the fundamental representation 
of these gauge groups can also be studied within the Dijkgraaf-Vafa setup
\cite{acfh1,mcg1,br1,suzuki1,reino1,ow1,fo1}.
For $U(N)$ gauge group, it was shown that the effective superpotential 
gets contributions from  the matrix model planar diagrams with
zero or one boundary \cite {acfh1,br1}. 
In \cite {reino1}, it has been shown that 
the $U(N)$ effective superpotential for
theories with $N_f$ fundamental flavors can be calculated in terms of
quantities computed in the pure gauge theory. 

Further, geometric engineering of these supersymmetric theories 
with $N_f$ fundamental flavors was considered in \cite {ook1} by 
placing $D5$ branes at locations given by the the mass 
$m_a$ ($a=1,2,\ldots N_f$) which are not the zeros of $W'(x)=0$. 
Though a formal expression for effective potential is derived, 
explicit form is given only for quadratic potential of the adjoint matter. 
The $SO/Sp$ effective superpotential computations from 
matrix model approach have been presented for tree level 
superpotential of adjoint matter upto quartic terms \cite {ow1,fo1}.
In this paper we consider ${\cal N}=1$ supersymmetric $SO(N_c)$ gauge theory
with arbitrary tree level superpotential of one chiral superfield in the
adjoint representation and $N_f$ fundamental matter multiplets.
We use the technique developed in \cite {reino1} to calculate the 
exact effective superpotential of this theory.

The organization of the paper as follows: In section 2, we briefly
discuss the relevant matrix model and its free energy. Then we
discuss the $SO(N_c)$ effective superpotential for ${\cal N}=1$
supersymmetric theory with fundamental matter in section 3.
In particular, we obtain a neat expression for the
effective superpotential for a most general tree level superpotential 
involving adjoint matter field. This is the main result of the 
paper. In section 4, we recapitulate the geometric considerations
of dualities. Then, we  evaluate the effective superpotential for 
a quartic tree level potential and show that the results agree with 
our expressions in section 3. We conclude with summary and 
discussions in section 5.

\section{Relevant Matrix Model}
Let us consider ${\cal N}=1$ supersymmetric $SO(N_c)$ gauge theory with one
adjoint field $\Phi$ and $N_f$
flavors of quarks $Q^I$'s with mass $m_I$'s ($I=1,2,\ldots N_f$) in the 
vector(fundamental) representation. The tree level superpotential of this
theory is given by \cite{ow1}
\begin{equation}
W_{tree}=W(\Phi)+\sum_{I=1}^{N_f}\left(\tilde{Q} \Phi Q + 
m_I \tilde{Q}Q\right)~, 
\end{equation} 
where $W(\Phi)$ is a polynomial with even powers of $\Phi$:
\begin{equation} 
W(\Phi) = \sum_{k=1}^{n+1} \frac{g_{2k}}{2k} Tr{\Phi}^{2k} \label{tree1}\, .
\end{equation}
According to Dijkgraaf-Vafa conjecture, the effective superpotential of
this theory can be obtained from the planar limit of the
matrix model whose tree level potential
is proportional to $W_{tree}$. Hence the partition function of the matrix
model is \cite{ow1}
\begin{equation}
Z = e^{-{\cal F}} = 
\int {\cal D}\Phi {\cal D}Q \, \,exp\left\{-\frac{1}{g_s}\left[W(\Phi)+
\sum_{I=1}^{N_f}\left(\tilde{Q} \Phi Q + m_I \tilde{Q}Q\right)\right]\right\}\, ,
\label {pf1}
\end{equation}
where $\Phi$ is $M\times M$ real antisymmetric matrix and $Q,\, \tilde{Q}$ are
$M$ dimensional vectors. For this theory, the Dijkgraaf-Vafa
conjecture can be generalized \cite {ronne1} to obtain effective superpotential
as a function of glueball field $S= Tr W_{\alpha}W^{\alpha}$:
\begin{equation}
W_{eff} = N_c \frac{\partial {\cal F}_{s^2}}{\partial S}
+ 4 {\cal F}_{RP^2} + {\cal F}_{D^2}\, , \label {weff2}
\end{equation}
where ${\cal F}_{s^2}$ is a free energy of a diagram with topology of
sphere, ${\cal F}_{RP^2}$ is free energy of the diagram with cross-cap one
(topology of $RP^2$) and ${\cal F}_{D^2}$ is the free energy of the
diagram with one boundary (topology of $D^2$).
It is also known that \cite {iho1}
\begin{equation}
{\cal F}_{RP^2}=-\frac{1}{2}
\frac{\partial {\cal F}_{s^2}}{\partial S}\, .
\end{equation}
Using this $W_{eff}$ becomes
\begin{eqnarray}
W_{eff} &=& (N_c-2) \frac{\partial {\cal F}_{s^2}}{\partial S}
+ {\cal F}_{D^2} \nonumber \\
&=& (N_c-2) \frac{\partial {\cal F}_{\chi=2}}{\partial S} + {\cal F}_{\chi=1}\nonumber \\
&=& W_{VY} + (N_c-2) \frac{\partial {\cal F}^{pert}_{\chi=2}}{\partial S} 
+ {\cal F}_{\chi=1}\,,
\label {weff3}
\end{eqnarray}
where $W_{VY}$ denotes the Veneziano-Yankielowicz potential \cite {vy1}.
Here we have absorbed ${\cal F}_{RP^2}$ in ${\cal F}_{\chi=2}$ and
${\cal F}_{\chi=1}$ contains the contribution to the free energy
coming from the fundamental matter.
As proposed by Dijkgraaf-Vafa, we need to take the planar limit of the
matrix model. The planar limit can be obtained by taking 
$(M,N_f \rightarrow \infty)$ as well as $g_s \rightarrow 0 $ such that
$S=g_s M$ and $S_f=g_s N_f$ are finite. 
The total free energy of this matrix model can be expressed as an expansion
in genus $g$ and the number of quark loops $h$: 
\begin{equation} 
{\cal F} = \sum_{g,h} g_s^{2g-2} {S_f}^h {\cal F}_{g,h}(S)~. \label {fe1}
\end{equation} 
Assuming that the fundamental quarks are massive compared to adjoint
matter, we can integrate out the fundamental matter 
fields appearing quadratically in the partition function (\ref{pf1})
to give 
\begin{equation}
Z=e^{-{\cal F}} = \int {\cal D}\Phi \, exp\left[-\frac{1}{g_s}
Tr\left( W(\Phi) + S_f \sum_{I=1}^{N_f} log(\Phi + m_I)\right)\right]~.
\label{fe2}
\end{equation} 
We are now in a position to calculate ${\cal F}_{\chi=1}$ and 
${\cal F}_{\chi=2}$ contributions to the superpotential.
In the following section we apply the 
method developed in \cite {reino1} for the $SO(N_c)$ gauge theory with 
one adjoint matter field and $N_f$ fundamental flavors.

\section{Effective Superpotential}
We wish to compute the exact effective superpotential 
of ${\cal N}=2 $ supersymmetric $SO(N_c)$ gauge theory 
with $N_f$ flavors of quark loops in the fundamental
representation, broken to ${\cal N}=1 $ by addition of a tree level
superpotential $W(\Phi)$ given by eqn.(\ref{tree1}). 
Following the arguments in \cite {reino1}, for the effective 
superpotential evaluation we can still look at a point in the quantum moduli 
space of ${\cal N}=2$ pure gauge theory where 
$r=[N_c/2]$ (rank of $SO(N_c)$) monopoles become 
massless \cite {jo1}. This corresponds to the point 
where the Seiberg-Witten curve factorizes completely.

We have seen that the effective superpotential
of this theory gets contributions from free energies ${\cal F}_{\chi=2}$ and 
${\cal F}_{\chi=1}$ of the matrix model described in the previous section.
We shall first calculate
the contribution to the superpotential coming from ${\cal F}_{\chi=2}$ 
using the moduli associated with Seiberg-Witten factorization.
\subsection{Contribution of ${\cal F}_{\chi=2}$}
In this section we compute the contribution of ${\cal F}_{\chi=2}$ to the
effective superpotential. This contains free energies of the diagrams having
topology of $S^2$ and $RP^2$. 
Taking derivative of eqn.(\ref{fe1}) with respect to $g_s$ 
\begin{equation} 
\frac{\partial{\cal F}}{\partial g_s} = \sum_{g,h} g_s^{2g-3}(S_f)^h \left(
(2g-2){\cal F}_{g,h} + S \frac{\partial {\cal F}_{g,h}}{\partial S} \right)
+\sum_{g,h} h g_s^{2g-3} (S_f)^h {\cal F}_{g,h}(S)\, . \label{dgsfe}
\end{equation}
According to Dijkgraaf-Vafa, one should take the planar limit on the
matrix model side. Also we take number of quark loops, $h=0$
for $\chi=2$ free-energy computation.
Planar limit of the above equation gives
\begin{equation} 
\frac{\partial{\cal F}}{\partial g_s} = g_s^{-3} \left(S \frac{\partial {\cal F}_{\chi=2}}
{\partial S} - 2 {\cal F}_{\chi=2}\right)\, .
\end{equation} 
We can also differentiate eqn.(\ref{fe2}) with respect to $g_s$ to give
\begin{equation} 
\frac{\partial{\cal F}}{\partial g_s} = -g_s^{-2} \langle TrW(\Phi) \rangle\, .
\end{equation}
From the above two equations 
\begin{equation} 
g_s \langle TrW(\Phi) \rangle = 2 {\cal F}_{\chi=2} 
- S \frac{\partial {\cal F}_{\chi=2}}{\partial S}\, . \label{f21}
\end{equation} 
The form of $W(\Phi)$ shows that the LHS contains the vacuum expectation
values $\langle Tr\Phi^{2p} \rangle$. It is clear from the above equation
that once we obtain the vevs $\langle Tr\Phi^{2p} \rangle$, we can easily
compute ${\cal F}_{\chi=2}$. In the case of ${\cal N}=2$ $SO(N_c)$ gauge
theory, the moduli are given by $u_{2p}=\frac{1}{2p}Tr\Phi^{2p}$. 
We are interested in the complete factorization
of the Seiberg-Witten curve. 
The moduli that factorizes the Seiberg-Witten curve are given by \cite{jo1}
\begin{equation} 
\langle u_{2p} \rangle  
 = \frac{N_c-2}{2p} C^p_{2p} \Lambda^{2p}\, ,
\end{equation}
where $C^i_j=\frac{j!}{i!(j-i)!}$ and
$\Lambda$ is the scale governing the running of the gauge coupling
constant. The matrix model calculation of the vevs of the moduli
done in the context of $SU(N_c)$ \cite {nsw1, nsw2} can be
extended to $SO(N_c)$ giving
\begin{equation}
\langle u_{2p} \rangle = (N_c-2) \frac{\partial}{\partial S} \frac{g_s}{2p}
\langle Tr {\Phi}^{2p} \rangle\, .
\end{equation}
It is obvious from the above two equations that
\begin{equation}
\frac{\partial}{\partial S} g_s \langle Tr {\Phi}^{2p} \rangle 
= C^p_{2p} \Lambda^{2p}~. \label{dtp1}
\end{equation} 
We denote the effective superpotential of pure $SO(N_c)$ gauge theory 
by $W_{eff}^0$. From eqn.(\ref{weff3}) we can write
\begin{equation} 
W_{eff}^0 = (N_c-2) \frac{\partial {\cal F}_{\chi=2}}{\partial S} = 
\frac{(N_c-2)}{2} S \left( -log \frac{S}{{\tilde \Lambda}^3} 
+1\right) + (N_c-2) \frac{\partial {\cal F}_{\chi=2}^{pert}}{\partial S}\, .
\label{weff4}
\end{equation} 
The first term in the above equation is the Veneziano-Yankeilowicz 
superpotential \cite {ahn1} and ${\tilde \Lambda}^{3(N_c-2)}$ is the
strong coupling scale of the ${\cal N}=1$ theory. The second term is 
perturbative in glueball superfield $S$ with
\begin{equation}
{\cal F}_{\chi=2}^{pert} = \sum_{n\geq 1} f_n^{\chi=2}(g_{2p}) S^{n+2}\, .
\label{fe3}
\end{equation}
Once we compute the functions $f_n^{\chi=2}(g_{2p})$, we will have
the exact superpotential of $SO(N_c)$ pure gauge theory. In order to
compute these functions we need to take the derivative of (\ref{f21})
with respect to $S$
\begin{equation}
\frac{\partial}{\partial S} g_s \langle TrW(\Phi)\rangle 
= \frac{\partial {\cal F}_{\chi=2}}
{\partial S} - S \frac{\partial^2 {\cal F}_{\chi=2}}{\partial S^2}\, .
\end{equation}
Substituting eqns.(\ref{weff4},\ref{fe3}) in the above equation, we get 
{\setlength \arraycolsep{2pt}
\begin{eqnarray}
(N_c-2) \frac{\partial}{\partial S} g_s \langle TrW(\Phi)\rangle &=& W_{eff}^0 
-S \frac{\partial W_{eff}^0}{\partial S} \nonumber \\\label{dtwp1}
&=& (N_c-2)\left[\frac{S}{2} 
- \sum_{n\geq 1} n(n+2) f_n^{\chi=2}(g_{2p}) S^{n+1}\right]\,. 
\end{eqnarray}
}

At the critical point of the superpotential that is when 
$\partial W^0_{eff}/\partial S = 0$ we have
\begin{equation} 
W_{eff}^0 = (N_c-2) \frac{\partial}{\partial S} g_s 
\langle Tr W(\Phi) \rangle 
= \sum_p g_{2p} \langle u_{2p} \rangle \,.
\end{equation}
The glueball superfield can be obtained at the critical point by the 
following relation \cite{jo1}:
\begin{equation}
S=\frac{\partial W_{eff}^0 }{\partial log \Lambda^{N_c-2}} 
= \sum_{p\geq 1} g_{2p} C^p_{2p} \Lambda^{2p}\,. \label{s1}
\end{equation}
Inserting eqn.(\ref{dtp1}) and eqn.(\ref{s1}) in eqn.(\ref{dtwp1}) one gets
\begin{equation}
\sum_{p\geq 1} \frac{1}{2p}g_{2p} C^p_{2p} \Lambda^{2p}
= \frac{1}{2} \sum_{p\geq 1} g_{2p} C^p_{2p} \Lambda^{2p}
-\sum_{n\geq 1} n(n+2) f_n^{\chi=2} (g_{2p}) S^{n+1}\,.
\end{equation}
Substituting glueball field $S$ in terms of $\Lambda$ (\ref {s1}) and
equating the powers of $\Lambda$ on both sides of the above
equation, we can extract the functions $f_n^{\chi=2} (g_{2p})$:  
\begin{eqnarray} 
f_1^{\chi=2} &=& \frac{1}{8} \frac{g_4}{g_2^2} \nonumber \\
f_{n\geq 2}^{\chi=2} &=& \frac{C^{n+1}_{2(n+1)}}{2^{n+2}(n+1)(n+2)}
\frac{g_{2(n+1)}}{g_2^{n+1}} \nonumber \\
&-& \sum_{l=1}^{n-1} \frac{l(l+2)}{n(n+2)}
f_l^{\chi=2} \sum_{{p_1,\ldots p_{l+1}=1} \atop{p_1+\ldots+p_{l+1}=n+1}}^{n+1}
\frac{C^{p_1}_{2p_1}g_{2p}\ldots C^{p_{l+1}}_{2p_{l+1}}g_{2p_{l+1}}}
{2^{n+1}g_2^{n+1}}\,, \label{ff2}
\end{eqnarray}
Now that we have computed the functions $f_n^{\chi=2} (g_{2p})$, the
$\chi=2$ contribution to the
effective superpotential of the $SO(N_c)$ theory with one adjoint
chiral superfield with arbitrary tree level superpotential is known 
exactly. From eqn.(\ref{weff4}) and eqn.(\ref{fe3}), it is given by
\begin{equation} 
W_{eff}^0  = (N_c-2)\left[\frac{S}{2} \left(-log \frac{S}{\tilde\Lambda^3}+1\right)
+ \sum_{n\geq 1}(n+2)f_n^{\chi=2} (g_{2p}) S^{n+1}\right]\,. \label{fw02}
\end{equation} 
In the case of quadratic tree level superpotential, that is when
$g_{2p}=0$ for $p\geq 2$, the functions $f_n^{\chi=2}(g_{2p})$ vanish for all $n$. 
And we can fix the coupling scale ${\tilde \Lambda}^3$ to $2g_2\Lambda^2$ 
by the requirement that $W^0_{eff}$ satisfies equation (\ref{s1}).
The $W_{eff}^0$ for the quartic tree level superpotential can be 
obtained by substituting $g_{2p}=0$ for $p\geq 3$
in the above result (\ref{fw02}): 
\begin{equation}
W_{eff}^0 = W_{VY} + (N_c-2)\left[\frac{3}{2}\left(\frac{g_4}{4g_2^2}\right)S^2
-\frac{9}{2}\left(\frac{g_4^2}{8g_2^4}\right)S^3
+\frac{45}{2}\left(\frac{g_4^3}{16g_2^6}\right)S^4+\ldots\right]\,.\label {fw0i}
\end{equation} 
This is in perfect agreement with the result of \cite{fo1} where it has
been evaluated in terms of the matrix model as well as IIB 
closed string theory on Calabi-Yau with fluxes.
We now compare the result (\ref{fw02}) with the corresponding result in the
$SU(N_c)$ gauge theory with one adjoint matter. The exact effective 
superpotential of $SU(N_c)$  theory has been obtained in \cite{reino1}. 
Comparison of the effective superpotentials of these two theories provides
the following equivalence:
\begin{equation}
W_{eff}^{0\,SO(N_c)} (g_{2p}) = \frac{N_c-2}{2N_c} \,
W_{eff}^{0\,SU(N_c)} (g^{\prime}_{2p}=2g_{2p})\,,
\end{equation}
which agrees with the relation obtained in \cite {jo1}.
We shall now address the fundamental matter contribution 
${\cal F}_{\chi=1}$ to the effective potential.

\subsection{Contribution of ${\cal F}_{\chi=1}$}

We differentiate the free energy given by eqn.(\ref {fe1}) 
with respect to $S_f$
\begin{equation} 
\frac{\partial {\cal F}}{\partial S_f} = \sum_{g,h} h g_s^{2g-2} (S_f)^{h-1}
{\cal F}_{g,h}(S)\,. \label{dfe1}
\end{equation} 
We are interested in genus $g=0$ and one quark loop $h=1$ contribution
in the planar limit $g_s\rightarrow 0$. 
The dominant term from 
eqn.(\ref{dfe1}) is ${\partial}_{S_f} {\cal F} = g_s^{-2} {\cal F}_{\chi=1}$.\\
Differentiation of eqn.(\ref{fe2}) with respect to $S_f$ gives
\begin{equation}
\frac{\partial {\cal F}}{\partial S_f} = g_s^{-1} \sum_{I=1}^{N_f} 
\langle Tr\, log(\Phi+m_I) \rangle\,, \label{dfe2}
\end{equation}  
This implies
\begin{equation}
{\cal F}_{\chi=1} = g_s \sum_{I=1}^{N_f} \langle Tr\, log(\Phi+m_I)\rangle\,.
\label{f11}
\end{equation}
Expanding the above equation around the critical point $\Phi=0$, we get
\begin{equation} 
{\cal F}_{\chi=1} = \sum_{I=1}^{N_f} \left(S\, log\, m_I  
- \sum_{k=1}^{\infty} \frac{(-1)^k}{k m_I^k} g_s\langle Tr{\Phi}^k\rangle\right)\,.
\end{equation}
Differentiating with respect to S and using eqn.(\ref{dtp1}) we get
\begin{equation}
\frac{\partial {\cal F}_{\chi=1}}{\partial S} = \sum_{I=1}^{N_f}
\left(log\, m_I - \sum_{k=1}^{\infty} \frac{1}{2\,k m_I^{2k}} C^k_{2k}\Lambda^{2k}\right)\,.
\end{equation}
Integrate above equation with respect to $S$
\begin{equation}
{\cal F}_{\chi=1} = \sum_{I=1}^{N_f}S\, log\, m_I 
- \sum_{I=1}^{N_f} \sum_{k,l\geq 1}\frac{l g_{2l} C^l_{2l} C^k_{2k}}
{2 k(k+l)m_I^{2k}} \Lambda^{2(k+l)} + A \,,\label{ff1}
\end{equation}
where $A$ is the constant of integration.
We postulate
\begin{equation}
A = \sum_{I=1}^{N_f} W_{tree}(m_I)\,,
\end{equation} 
and we will see in the next section that the result agrees with the one 
obtained from Calabi-Yau geometry with fluxes.
In order to write this expression in powers of $S$, we write 
${\cal F}_{\chi=1}$ as 
\begin{equation} 
{\cal F}_{\chi=1} = S \sum_{I=1}^{N_f} log\, m_I 
+ \sum_{n\geq 1} f_n^{\chi=1} (g_{2p}) S^{n+1} + \sum_{I=1}^{N_f} W_{tree}(m_I)\,.\label{fp1}
\end{equation}
Comparison with eqn.(\ref{ff1}) gives the following recursive relation for
the coefficients $f_n^{\chi=1} (g_{2p})$
\begin{eqnarray} 
f_1^{\chi=1} &=& - \frac{1}{4} \sum_{I=1}^{N_f} \frac{1}{m_I^2 g_2}\nonumber \\
f_{n\geq 1}^{\chi=1} &=& -\frac{1}{2^{n+1} g_2^{n+1}} \left( \sum_{I=1}^{N_f}
\sum_{k,l=1}^n \frac{l g_{2l} C^l_{2l} C^k_{2k}}{2 k(n+1)m_I^{2k}}\right.\nonumber \\
&+&\left. \sum_{q=1}^{n-1} f_q^{\chi=1} \sum_{{p_1,\ldots p_{q+1}} 
\atop{p_1+\ldots+p_{q+1}=n+1}}^{n+1} C^{p_1}_{2p_1} g_{2p_1}\ldots 
C^{p_{q+1}}_{2p_{q+1}} g_{2p_{q+1}} \right)\,. \label{ff12}
\end{eqnarray}
The eqn.(\ref{fp1}) alongwith eqn.(\ref{ff12}) gives the effective 
superpotential from fundamental matter, for the most general $W_{tree}$.
If we substitute $g_{2p}=0$ for $p \geq 2$ in the above result, we get
${\cal F}_{\chi=1}$ for the gauge theory with quadratic superpotential.
It is explicitly given by,
\begin{equation}
{\cal F}_{\chi=1} = \sum_{I=1}^{N_f} \left[S log\, m_I 
-\frac{1}{4} \frac{S^2}{{m_I}^2 g_2} 
- \frac{1}{8} \frac{S^3}{{m_I}^4 {g_2}^2}
- \frac{5}{48} \frac{S^4}{{m_I}^6 {g_2}^3} - \ldots \right] +A \,.\label {qua0}
\end{equation} 
Also substituting $g_{2p}=0$ for $p \geq 3$, we get ${\cal F}_{\chi=1}$ 
for the theory with quartic superpotential.
\begin{eqnarray} 
{\cal F}_{\chi=1} &=& \sum_{I=1}^{N_f} \left[S log\, m_I +
\left(- \frac{1}{4{m_I}^2 g_2}\right)S^2
+ \left(- \frac{1}{8{m_I}^4 {g_2}^2} + \frac{g_4}{4{m_I}^2 {g_2}^3}\right)S^3 \right.\nonumber \\
&+& \left. \left(-\frac{5}{48{m_I}^6 {g_2}^3} + \frac{9g}{32{m_I}^4 {g_2}^4}
- \frac{9 g^2}{16{m_I}^2 {g_2}^5}\right)S^4 + \ldots \right] + A \,.\label {quar1}
\end{eqnarray}
The total effective superpotential of the theory under consideration is
\begin{eqnarray}
W_{eff}&=&W^0_{eff} + {\cal F}_{\chi=1} \nonumber \\
&=& (N_c-2)\left[\frac{S}{2} \left(-log \frac{S}{2 g_2 \Lambda^2}+1\right)
+ \sum_{n\geq 1}(n+2)f_n^{\chi=2} (g_{2p}) S^{n+1}\right] \nonumber \\
&+& S \sum_{I=1}^{N_f} log\, m_I
+ \sum_{n\geq 1} f_n^{\chi=1} (g_{2p}) S^{n+1} + \sum_{I=1}^{N_f} W_{tree}(m_I)
\label{tweff}
\end{eqnarray}
It is important to realize the power of assimilating
Dijkgraaf-Vafa conjecture and the connections to factorization
of Seiberg-Witten curves. As a consequence, we have obtained a 
concise expression for $SO(N_c)$ effective
superpotential for arbitrary polynomials of tree level
superpotential. In order to make sure that the results
are consistent, we need to compare with other approaches.

In the next section, we compare the results with explicit answers  
obtained from geometric approach of dualities for tree level 
superpotential upto quartic term.

\section{ Geometric Engineering and Effective $SO$ Superpotential}
We will briefly recapitulate geometric dualities leading to the
computation of $SO$ superpotential.
\subsection{Geometric Transition}
Consider type IIB String theory compactified on an orientifold
of a resolved Calabi-Yau geometry whose singular limit is 
given by eqn.(\ref {singu}). For description of $SO$ gauge group,
$W(x)$ (\ref {singu}) must be even functions of $x$. 
Further, $W'(x)=0$ determines the
eigenvalues of $\Phi$ which can be $0, \pm i a_i's$. 

We are interested in ${\cal N}=1$ $SO(N_c)$ supersymmetric 
gauge theory in four dimensions. This can be realized by 
wrapping $N_c$ D5branes on ${\bf RP}^2$ of the 
orientifolded resolved geometry- i.e., we place all the $N_c$ 
branes at $x=0$ where eigenvalues of $\Phi$ are zero.
Invoking large $N$ duality \cite{vafa1,civ1, eot1}, 
the supersymmetric gauge theory is dual to IIB string 
theory on a deformed Calabi-Yau geometry with fluxes.
The deformed geometry is described by
\begin{equation}
k\equiv W'(x)^2+ f_{2n-2}(x)+ y^2+z^2 + w^2=0 \,,\label {defm}
\end{equation}
where $f_{2n-2}(x)$ is a $n-1$ degree polynomial in $x^2$. 
The three-cycles in this geometry can be given in terms of
basis cycles $A_i,B_i \in H_3(M, {\bf Z})$ ($i=1,2,\ldots 2n+1)$
satisfying symplectic pairing
$$(A_i, B_j)= -(B_j,A_i)= \delta_{ij}~,~~(A_i,A_j)=(B_i,B_j)=0~.$$
Here the pairing $(A,B)$ of three-cycles $A,B$ is defined as the
intersection number. For the deformed Calabi-Yau (\ref {defm}), these
three-cycles are constructed as ${\bf P}^1$ fibration
over the line segments between two critical points 
$x=0^+, 0^-, \pm i a_i^+, \pm i a_i^- \ldots$ 
of $W'(x)^2+f_{2n-2}(x)$ in $x$-plane. In particular,
$A_0$ cycle corresponds to ${\bf P}^1$ fibration over the
line segment $0^-< x < 0^+$ and $A_i$'s to be fibration over
the line segments $ia_i^-< x < ia_i^+$. The three-cycles
$B_0(B_i's)$  are non-compact and are given be fibrations over line
segments between $0 < x <\Lambda_0 (ia_i^+<x< i \Lambda_0)$
where $\Lambda_0$ is a cut-off. The deformed geometry (\ref {defm})
has ${\bf Z}_2$ symmetry and hence we can restrict the
discussion to the upper half of $x$-plane.
The holomorphic three-form $\Omega$ for the deformed geometry (\ref {defm})
is give by
\begin{equation}
\Omega = 2 {dx \wedge dy \wedge dz \over \partial k/\partial \omega}~.
\end{equation}
The periods $S_i$ and the dual periods $\Pi_i$ for this deformed geometry are 
$$S_i = \int_{A_i} \Omega~,~~ \Pi_i= \int_{B_i} \Omega~.$$
The dual periods in terms of prepotential ${\cal F}(S_i)$ is 
$\Pi_i= \partial {\cal F}/\partial S_i$. Using the fact that these
three cycles can be seen as ${\bf P}^1$ fibrations over appropriate
segments in the $x$-plane, the periods can be rewritten as 
integral over a one-form $\omega$ in the $x$-plane.
That is, $S_0= 1/(2 \pi i) \int_{0^-}^{0^+} \omega~, ~~   
\Pi_0= 1/(2 \pi i) \int_{0^+}^{\Lambda_0} \omega~, \ldots$
where the one-form $\omega$ is given by
\begin{equation}
\omega = 2 dx \left( W'(x)^2+ f_{2n-2}(x) \right)^{1 \over 2}~.
\end{equation}
The effective superpotential $W_{eff}^0$ (recall the suffix $0$ 
denotes the contribution from adjoint matter field $\Phi$) 
can be obtained as follows
\begin{equation}
-{1 \over 2 \pi i} W_{eff}^0= \int \Omega \wedge (H_R+ \tau H_{NS})\,,
\end{equation}
where $\tau$ is the complexified coupling constant of type IIB strings,
the $H_R$ and $H_{NS}$ denotes the RR-three form and NS-NS
three-form field strengths and their fluxes satisfy the following relations
under the geometric transition:
\begin{equation}
N_c-2 = \int_{A_0} H_R~,~~ \alpha = \int_{B_i} \tau H_{NS}~.
\end{equation}
For the classical solution $\Phi=0$ with $N$ D5-branes
at $x=0$, the dual theory will require the above $RR$-flux  
over $A_0$ cycle alone and a non-zero period $S_0 \equiv S$.
The prepotential ${\cal F}(S)$ will be the $\chi=2$ 
part of the matrix model free energy. 

Inclusion of matter in fundamental representations
in the geometric  framework corresponds to placing D5 branes 
at locations $x= m_a$ where $m_a$'s are the masses of 
$N_f$ fundamental flavors. These locations are not the zeros
of $W'(x)=0$. The fundamental matter contribution 
to the effective potential is given by \cite {ook1}:
\begin{equation}
W_{eff}^{flav}\equiv {\cal F}_{\chi=1}={1 \over 2} \sum_{a=1}^{N_f} 
\int_{m_a}^{\Lambda_0} \omega~. \label {flav}
\end{equation}
It is important to work out explicitly these formal integrals
for specific potentials and compare with our closed form expression
obtained for arbitrary potentials in section 3.

\subsection{Effective Superpotential for Quartic Potential}
In this subsection we consider the ${\cal N}=1$ $SO(N_c)$ gauge theory with
fundamental matter and quartic tree level superpotential:
\begin{equation}
W_{tree}(\Phi) = \frac{M}{2}Tr\Phi^2 + \frac{g}{4}Tr\Phi^4\,.
\end{equation} 
The geometry corresponding to this gauge theory is given by
\begin{equation} 
W^{\prime}(x)^2+f_2(x)+y^2+z^2+w^2=0\,,
\end{equation} 
where $f_2(x)$ is an even polynomial of degree 2.
We concentrate on the special classical vacuum $\Phi=0$, which 
is sometimes called as one cut solution in the
context of matrix models \cite {fo1}. We require 
the critical points of $W'(x)^2+f_2(x)$ to be $0^+, 0^-$.
This is achieved by the following one form:
\begin{equation} 
\omega = 2\sqrt{W^{\prime}(x)^2+f(x)}dx = 2g(x^2+\Delta+2\mu^2)
\sqrt{(x-2\mu)(x+2\mu)}~, \label{omega}
\end{equation}
where $0^{\pm}= \pm 2 \mu$ and $\Delta = \frac{M}{g}$.
The period integral can be computed from
\begin{equation} 
S = \frac{1}{2\pi i} \int_{-2\mu}^{2\mu} \omega dx~.
\end{equation} 
For the quartic $W_{tree}$, it is explicitly given by
\begin{equation} 
S = 2g\mu^2(\Delta+3\mu^2)~.
\end{equation} 
Equivalently, the above equation is quadratic in $\mu^2$ 
which can be solved to give the roots. Discarding
the negative root, we take the other root 
\begin{equation} 
\mu^2 = -\frac{\Delta}{6} + \frac{\Delta}{6} \sqrt{1+\frac{6S}{g\Delta^2}}~.
\end{equation} 
In fact, it is this part of the computation which prevents generalization
to potentials of powers higher than $\Phi^4$. 
For the given one-form, the $\chi=2$ contribution to the free-energy
is
\begin{equation}
W_{eff}^0= (N_c -2) {\partial \cal F (S) \over \partial S}= (N_c-2)
\int_{2 \mu}^{\Lambda^0} \omega dx~.
\end{equation}
This result in the $\Lambda_0 \rightarrow  \infty$ limit
agrees with eqn. (\ref {fw0i}).

The effective superpotential that comes from the contribution of flavors 
(\ref {flav}) is
{\setlength \arraycolsep{2pt}
\begin{eqnarray} 
W_{eff}^{flavor} &=& - g \sum_{I=1}^{N_f} \left[m_I \frac{\Delta+3\mu^2}{2}\sqrt{{m_I}^2-4\mu^2}
+ \frac{m_I}{4}\left({m_I}^2-4\mu^2\right)^{3/2}
+ \frac{\mu^2}{2}\left(2\Delta+3\mu^2\right) \right.\nonumber \\
&+& \left. 2 \mu^2\left(\Delta+3\mu^2\right)log(2\Lambda_0)
- 2 \mu^2\left(\Delta+3\mu^2\right)log\left(m_I+\sqrt{{m_I}^2-4\mu^2}\right)\right]\,.
\end{eqnarray} 
}
In obtaining the above result, we take the limit $\Lambda_0 \rightarrow \infty$
and ignore the  $\Lambda_0$ dependent terms. Substituting $\mu^2$ and
rewriting in powers of $S$ agrees with the our expansion (\ref {quar1}).

If we take $g \rightarrow 0$ limit in the above equation, we get
$W_{eff}^{flavor}$ of the $SO(N_c)$ gauge theory with quadratic tree level
superpotential. 
\small{
\begin{equation} 
W_{eff}^{flavor} = - \sum_{I=1}^{N_f} \left[\frac{S}{2} + \frac{M {m_I}^2}{2}
\sqrt{1-\frac{2S}{M {m_I}^2}} + S log\left(\frac{\Lambda_0}{m_I}\right)
- S log\left(\frac{1}{2}+\frac{1}{2}\sqrt{1-\frac{2S}{M {m_I}^2}}\right)\right]
\,.
\end{equation}}
If we replace $M$ by $M'/2$, we get the Affleck-Dine-Seiberg
$SU(N)$ superpotential \cite {ads1}. Expanding the above
equation in powers of $S$ agrees with eqn. (\ref {qua0}).

\section{Summary and Discussion}
In this paper, we have derived $SO(N_c)$ effective superpotential
for the supersymmetric theory with $N_f$ fundamental flavors (\ref{tweff}).
Using Dijkgraaf-Vafa conjecture and also the Sieberg-Witten factorization,
we have obtained the effective superpotential  for a most general
tree level potential $W_{tree}(\Phi^2)$.
We have shown agreement with the results from  the geometric considerations
of superstring dualities for a tree level potential upto quartic terms.
                                                                                
Though we have concentrated on the $SO$ gauge group, it appears
that the fundamental matter contribution to the $Sp$ (symplectic)
effective superpotential will be identical (${\cal F}_{\chi=1}$).
The effective potential in the absence of matter is well-studied
from various approaches which leads to the replacement of factor $N_c-2$
in eqn.(\ref{tweff}) by $N_c+2$ to get $W_{eff}^0$ for $Sp(N_c)$ gauge group.
                                                                                
Within supersymmetric theories, the effective superpotentials for
different regimes like $N_f=N_c$ or $N_f<N_c$ or $N_f > N_c$ could
be addressed. It is still a challenging problem to see such
distinction within the matrix model approach.

\newpage
\noindent
{\bf Acknowledgments}
                                                                                
\noindent
PB would like to thank CSIR for the grant.
The work of PR is supported by Department of Science and Technology
grant under `` SERC FAST TRACK Scheme for Young Scientists''.
\bibliographystyle{JHEP}
\bibliography{ref4}

\providecommand{\href}[2]{#2}\begingroup\raggedright\begin{thebibliography}{10}

\bibitem{gv1}
R.~Gopakumar and C.~Vafa, {\it On the gauge theory/geometry correspondence},
  {\em Adv. Theor. Math. Phys.} {\bf 3} (1999) 1415--1443. {hep-th/9811131}.

\bibitem{vafa1}
C.~Vafa, {\it Superstrings and topological strings at large n},  {\em J. Math.
  Phys.} {\bf 42} (2001) 2798--2817. {hep-th/0008142}.

\bibitem{civ1}
F.~Cachazo, K.~Intriligator, and C.~Vafa, {\it A large n duality via a
  geometric transition},  {\em Nucl. Phys. B} {\bf 603} (2001) 3--41.
  {hep-th/0103067}.

\bibitem{dv1}
R.~Dijkgraaf and C.~Vafa, {\it Matrix models, topological strings and
  supersymmetric gauge theories},  {\em Nucl. Phys. B} {\bf 644} (2002) 3--20.
  {hep-th/0206255}.

\bibitem{dv2}
R.~Dijkgraaf and C.~Vafa, {\it On geometry and matrix models},  {\em Nucl.
  Phys. B} {\bf 644} (2002) 21--39. {hep-th/0207106}.

\bibitem{dv3}
R.~Dijkgraaf and C.~Vafa, {\it A perturbative window into non-perturbative
  physics},  {\em hep-th/0208048}.

\bibitem{ferr1}
F.~Ferrari, {\it On exact superpotentials in confining vacua},  {\em Nucl.
  Phys. B} {\bf 648} (2003) 161--173. {hep-th/0210135}.

\bibitem{dglvz1}
R.~Dijkgraaf, M.~T. Grisaru, C.~Lam, C.~Vafa, and D.~Zanon, {\it Perturbative
  computation of glueball superpotentials},  {\em Phys. Lett. B} {\bf 573}
  (2003) 138--146. {hep-th/0211017}.

\bibitem{cdsw1}
F.~Cachazo, M.~Douglas, N.~Seiberg, and E.~Witten, {\it Chiral rings and
  anomalies in supersymmetric gauge theory},  {\em JHEP} {\bf 0212} (2002) 071.
  {hep-th/0211170}.

\bibitem{sv1}
S.~Sinha and C.~Vafa, {\it So and sp chern-simons at large n}, .
  {hep-th/0012136}.

\bibitem{eot1}
J.~Edelstein, K.~Oh, and R.~Tatar, {\it Orientifold, geometric transition and
  large $n$ duality for $so/sp$ gauge theories},  {\em JHEP} {\bf 0105} (2001)
  009. {hep-th/0104037}.

\bibitem{iho1}
H.~Ita, H.~Hieder, and Y.~Oz, {\it Perturbative computation of glueball
  superpotentials for so(n) and usp(n)},  {\em JHEP} {\bf 01} (2003) 018.
  {hep-th/0211261}.

\bibitem{achkr1}
S.~K. Ashok, R.~Corrado, N.~Halmagyi, K.~Kennaway, and C.~Romelsberger, {\it
  Unoriented strings, loop equations, and n=1 superpotentials from matrix
  models},  {\em Phys. Rev. D} {\bf 67} (2003) 086004. {hep-th/0211291}.

\bibitem{jo1}
R.~Janik and N.~Obers, {\it So(n) superpotential, seiberg-witten curves and
  loop equations},  {\em Phys. Lett. B} {\bf 553} (2003) 309--316.
  {hep-th/0212069}.

\bibitem{ikrsv1}
K.~Intriligator, P.~Kraus, A.~V. Ryzhov, M.~Shigemori, and C.~Vafa, {\it On low
  rank classical groups in string theory, gauge theory and matrix models},
  {\em Nucl. Phys. B} {\bf 682} (2004) 45--82. {hep-th/0311181}.

\bibitem{argurio1}
R.~Argurio, {\it Effective superpotential for $u(n)$ with antisymmetric
  matter},  {\em JHEP} {\bf 0409} (2004) 031. {hep-th/0406253}.

\bibitem{ac1}
L.~F. Alday and M.~Cirafici, {\it Effective superpotentials via konishi
  anomaly},  {\em JHEP} {\bf 0305} (2003) 041. {hep-th/0304119}.

\bibitem{acfh1}
R.~Argurio, V.~Campos, G.~Ferretti, and R.~Heise, {\it Exact superpotentials
  for theories with flavors via matrix integral},  {\em Phys. Rev. D} {\bf 67}
  (2003) 065005. {hep-th/0210291}.

\bibitem{mcg1}
J.~McGreevy, {\it Adding flavor to dijkgraaf-vafa},  {\em JHEP} {\bf 01} (2003)
  047. {hep-th/0211009}.

\bibitem{br1}
I.~Bena and R.~R, {\it Exact superpotentials in n=1 theories with flavor and
  their matrix model formulation},  {\em Phys. Lett. B} {\bf 555} (2003)
  117--125. {hep-th/0211075}.

\bibitem{suzuki1}
H.~Suzuki, {\it Perturbative derivation of exact superpotential for meson
  fields from matrix theories with one flavor},  {\em JHEP} {\bf 0303} (2003)
  005. {hep-th/0211052}.

\bibitem{reino1}
M.~Gomez-Reino, {\it Exact superpotentials, theories with flavor and confining
  vacua},  {\em JHEP} {\bf 0406} (2004) 051. {hep-th/0405242}.

\bibitem{ow1}
Y.~Ookouchi and Y.~Watabiki, {\it Effective superpotentials for so/sp with
  flavor from matrix models},  {\em Mod. Phys. Lett. A} {\bf 18} (2003)
  1113--1126. {hep-th/0301226}.

\bibitem{fo1}
H.~Fuji and Y.~Ookouchi, {\it Comments on effective superpotentials via matrix
  models},  {\em JHEP} {\bf 0212} (2002) 067. {hep-th/0210148}.

\bibitem{ook1}
Y.~Ookouchi, {\it N=1 gauge theory with flavor from fluxes},  {\em JHEP} {\bf
  01} (2004) 014. {hep-th/0211287}.

\bibitem{ronne1}
P.~B. R{\o}nne, {\it On the dijkgraaf-vafa conjecture},  {\em hep-th/0408103}.

\bibitem{vy1}
G.~Veneziano and S.~Yankeilowicz, {\it An effective lagrangian for the pure n=1
  supersymmetric yang-mills theory},  {\em Phys. Lett. B} {\bf 113} (1982) 231.

\bibitem{nsw1}
S.~Naculich, H.~Schnitzer, and N.~Wyllard, {\it The n=2 u(n) gauge theory
  prepotential and periods from a perturbative matrix model calculation},  {\em
  Nucl. Phys. B} {\bf 651} (2003) 106--124. {hep-th/0211123}.

\bibitem{nsw2}
S.~Naculich, H.~Schnitzer, and N.~Wyllard, {\it Matrix model approach to the
  n=2 u(n) gauge theory with matter in fundamental representation},  {\em JHEP}
  {\bf 0301} (2003) 015. {hep-th/0211254}.

\bibitem{ahn1}
C.~h.~Ahn, {\it Supersymmetric so(n)/sp(n) gauge theory from matrix model:
  Exact mesonic vacua},  {\em Phys. Lett. B} {\bf 560} (2003) 116--127.
  {hep-th/0301011}.

\bibitem{ads1}
I.~Affleck, M.~Dine, and N.~Seiberg, {\it Dynamical supersymmetry breaking in
  supersymmetric qcd},  {\em Nucl. Phys. B} {\bf 241} (1984) 493--534.

\end{thebibliography}\endgroup
\end{document}